\documentclass[prd,amsmath,amssymb,nofootinbib,superscriptaddress,showpacs]{revtex4}

\usepackage{ams math}
\usepackage{ams fonts}
\usepackage{graphics}
\usepackage{subfigure}

\newcommand{\scal}[2]{\vec #1\cdot\vec #2}

\begin{document}
\title{Planck-scale Modifications to Electrodynamics\\
 characterized by a space-like symmetry-breaking vector}
\author{Giulia GUBITOSI}
\affiliation{Dipartimento di Fisica, Universit\`a di Roma ``La
Sapienza'', \\ and Sezione Roma1 INFN\\P.le Aldo Moro 2, 00185 Roma, Italy}
\author{ Giuseppe GENOVESE}
\affiliation{Dipartimento di Matematica, Universit\`a di Roma ``La
Sapienza'', \\ and Sezione Roma1 INFN\\P.le Aldo Moro 2, 00185 Roma, Italy}
\author{Giovanni AMELINO-CAMELIA}
\affiliation{Dipartimento di Fisica, Universit\`a di Roma ``La
Sapienza'', \\ and Sezione Roma1 INFN\\P.le Aldo Moro 2, 00185 Roma, Italy}
\author{Alessandro MELCHIORRI}
\affiliation{Dipartimento di Fisica, Universit\`a di Roma ``La
Sapienza'', \\ and Sezione Roma1 INFN\\P.le Aldo Moro 2, 00185 Roma, Italy}


\begin{abstract}

In the study
of Planck-scale (``quantum-gravity induced") violations of Lorentz symmetry,
an important role was played by the deformed-electrodynamics
model introduced by Myers and Pospelov.
Its reliance on conventional effective quantum field theory,
and its description of symmetry-violation effects simply in terms of
a four-vector with nonzero component only in the time-direction,
rendered it an ideal target for experimentalists and a natural
concept-testing ground
 for many theorists.
At this point however the experimental limits
on the single Myers-Pospelov parameter, after improving steadily over these
past few years, are ``super-Planckian", {\it i.e.} they take the model
out of actual interest from a conventional quantum-gravity perspective.
In light of this we here argue that it may be appropriate to move on to
the next level of complexity, still with vectorial symmetry violation
but adopting a generic four-vector. We also offer a preliminary characterization
of the phenomenology of this more general framework, sufficient to expose
a rather significant increase in complexity with respect to the original
Myers-Pospelov setup. Most of these novel features are linked to
the presence of spatial anisotropy, which is particularly pronounced when the symmetry-breaking
vector is space-like, and they are such
that they reduce the bound-setting power of certain types of observations
in astrophysics.

\end{abstract}

\maketitle

\baselineskip 13pt

\section{Introduction}
A large effort has been
devoted over the last decade
(see, {\it e.g.},
Refs.~\cite{grbgac,LQGDispRel,astroSchaefer,astroBiller,astroKifune,gacNature1999,urrutiaPRL,gacPIRANprd,jaconature,PiranNeutriNat,ellisPLB2009,gacSMOLINprd2009,fermiNATURE,gacPRL2009}
and references therein)
toward establishing that it is possible to
actually study experimentally some
minute effects
introduced at the
ultra-high ``Planck scale" $M_P (\simeq 1.2 \cdot 10^{28} eV$),
the scale expected to characterize quantum-gravity effects.
At this point the scopes of this ``quantum-gravity phenomenology"~\cite{gacLRR}
extend over a rather large ensemble of candidate quantum-gravity
effects, inspired by (and/or formalized within) several models that
are believed to be relevant for the understanding of the quantum-gravity problem.
We here focus on one of these research programmes which has been driven
by a model first introduced by
Myers and Pospelov~\cite{Myers2003},
as a candidate description of the Lorentz-symmetry-violation effects
that are expected in some approaches to the quantum-gravity
problem~\cite{grbgac,LQGDispRel,urrutiaPRL,gacSMOLINprd2009}.
This model adopts effective field theory for the description
of Lorentz-symmetry-violation effects that are suppressed by a single power of the Planck scale
(linear in $1/M_P$) and its proposal was primarily grounded on the observation~\cite{Myers2003}
that
there is a unique such correction term
which could be added to Maxwell theory,
\begin{equation}
 \Delta \mathcal L_{QG}= \frac{1}{2M_P} n^\alpha F_{\alpha\delta}n^\sigma
  \partial_\sigma(n_\beta\varepsilon^{\beta\delta\gamma\lambda}F_{\gamma\lambda})
  ~, \label{eq:lagrangian}
\end{equation}
if one enforces some relatively weak assumptions, including gauge invariance
and the characterization of the symmetry-breaking structure in terms
of an external four-vector $n^\alpha$.

Myers and Pospelov provided an even more definite and manageable
framework by restricting their attention~\cite{Myers2003} to the case in which
the four-vector $n_\alpha$
only has a time component, $n_\alpha = (n_0,0,0,0)$.
Then, upon introducing the convenient notation  $\xi \equiv (n_0)^3$,
one arrives at the following modified Maxwell Lagrangian density:
\begin{equation}
 \mathcal L_{MP}=-\frac{1}{4}F_{\mu\nu}F^{\mu\nu}+\frac{\xi}{2M_{P} }
  \varepsilon^{jkl} F_{0 j} \partial_0F_{k l}\, ,
 \label{eq:MP}
\end{equation}
and in particular it is then possible to exploit the simplifications provided
by spatial isotropy.
This Myers-Pospelov effective-field-theory model of Planck-scale modified
electromagnetism has attracted much attention over the last few years.
For phenomenologists it provided an ideal target
(see {\it e.g.} Refs~\cite{gacLRR,mattinglyLRR,liberati0805,Galaverni:2007tq, Maccione:2008iw} and references
therein), because of the presence of a single parameter
and because (unlike most other fashionable proposals for the study of the quantum-gravity
problem~\cite{gacLRR}) its reliance on standard effective field theory poses no challenges
at the level of ``physical interpretation" of the formalism.

This vigorous effort of investigation of the Myers-Pospelov model has produced
a quick pace of improvement of experimental bounds, and, while the rough
estimate invited by a quantum-gravity intuition~\cite{gacLRR,mattinglyLRR,liberati0805}
would be $\xi \sim 1$,
the Myers-Pospelov parameter $\xi$ is now constrained to be much smaller than 1,
with some analyses~\cite{Galaverni:2007tq, Maccione:2008iw}
even suggesting a bound at the level $\xi < 10^{- 15}$.
We here observe that however these bounds are not applicable to the general
correction term $ \Delta \mathcal L_{QG}$ of Eq.~(\ref{eq:lagrangian}), since they exploit
significantly the spatial isotropy regained by the {\it ad hoc}
choice $n_\alpha = (n_0,0,0,0)$.
And actually this {\it ad hoc} choice is only available for a
restricted class of frames of reference:
even imposing ``by brute force" $n_\alpha = (n_0,0,0,0)$  in some desired frame of reference,
then the four-vector $n_\alpha$ will of course still acquire a spatial component
in other (boosted) frames. Since the main strategy for constraining
the Myers-Pospelov parameter has relied on various astrophysics observations,
conducted in different ``laboratory frames", these are concerns that necessarily must be
investigated, at least in order to establish to which extent those limits
are vulnerable to the presence of a (perhaps small, but necessarily nonzero)
spatial component in frames other than the ``preferred frame".

In the next section we therefore propose a phenomenology centered on the more
general form of the $ \Delta \mathcal L_{QG}$ of Eq.~(\ref{eq:lagrangian}),
involving therefore an arbitrary (four-parameter) four-vector $n_\alpha$,
and we describe the resulting equations of motion for the electromagnetic field.
Since the types of data that are most useful and are likely to still be most useful to
set bounds on this framework concern regimes that involve classical electromagnetic waves,
we shall here be satisfied with an analysis confined at the level of some modified
Maxwell equation for classical electromagnetic waves. In this respect we adopt
the same perspective
 of the original analysis by
Myers and Pospelov~\cite{Myers2003}, but for our purposes it is valuable to
provide, as we shall, a more detailed description of the Planck-scale modifications
of classical electromagnetic waves, whereas Ref.~\cite{Myers2003} focused
exclusively on the form of the dispersion (/``on-shell") relation.

In Section~3 we investigate the features that are likely to be most
relevant from the phenomenology perspective, which concern
dispersion, birefringence and a possible longitudinal component.
In Section~4 we provide a rough quantitative characterization of the
effects introduced by the spatial components of $n_\alpha$,
focusing mainly on cases with space-like symmetry-breaking vector
and stressing
that the magnitude of the effects is not exclusively governed by the magnitude
of the spatial components of $n_\alpha$: there are direction-dependent (anisotropic)
effects, and even small values of the spatial components of $n_\alpha$
produce large effects within a certain range of directions.
Section~5 offers some closing remarks.

\section{Modified Maxwell equations and analogy with anisotropic media}
By adding the Planck-scale correction term (\ref{eq:lagrangian}) to Maxwell's Lagriangian
we arrive at a modified Lagrangian density for electrodynamics of the form
\begin{equation}
\mathcal L_{QG}=-\frac{1}{4}F_{\mu\nu}F^{\mu\nu}
 +\frac{1}{2M_P} n^\alpha F_{\alpha\delta}n^\sigma
  \partial_\sigma(n_\beta\varepsilon^{\beta\delta\gamma\lambda}F_{\gamma\lambda})
  ~, \label{eq:lagrangianJOC}
\end{equation}
from which one easily derives the associated   modified Maxwell equations:
\begin{eqnarray}
0&=&\vec \nabla \times \vec B-\frac{\partial\vec E}{\partial t} -\frac{2 }{M_{P}}  (n_0 \frac{\partial}{\partial t}-\vec n\cdot \vec\nabla)^2  (\vec n\times \vec E+n_0 \vec B)\\
0&=&\vec \nabla\cdot \vec E+\frac{2}{M_{P}}  (n_0 \frac{\partial}{\partial t}-\vec n\cdot \vec\nabla)^2\,\vec n\cdot \vec B\\
0&=&  \vec\nabla \times\vec E+\frac{\partial \vec B}{\partial t}\\
0&=&\vec\nabla \cdot\vec  B
\end{eqnarray}

For the case of plane waves, in which we are primarily interested,
these modified Maxwell equations take the form
\begin{eqnarray}
 \vec k \times \vec B &=& - \omega \vec E- i \frac{2 }{M_{P}}   ( n_0 \omega+\vec n\cdot \vec k)^2  (\vec n\times \vec E+n_0 \vec B)\nonumber\\
\vec k \times\vec E&=&  \omega \vec B\nonumber\\
\vec k\cdot \vec E &=&-i \frac{2}{M_{P}}  ( n_0 \omega+\vec n\cdot \vec k)^2\,\vec n\cdot \vec B\nonumber\\
\vec k \cdot\vec  B&=&0.
\label{eq:maxwell}
\end{eqnarray}
Interestingly
these equations are rather similar to the ones
that govern ordinary propagation of electromagnetic radiation in
certain anisotropic media~\cite{bornwolf}.
In particular, one could view (\ref{eq:maxwell})
as equations of propagation in a material with polarization vector
\begin{equation}
\vec P= \frac{2}{M_{P}} \omega \left|\left(\vec{n}
+n_0 \frac{\vec{k}}{\omega }\right)\right|\left(n_0
+\vec{n}\cdot \frac{\vec{k}}{\omega }\right)^2\;(- i \hat v)\times \vec{E},
\end{equation}
where we
introduced the notation
$$\hat v=\frac{\vec{n}+n_0 \frac{\vec{k}}{\omega }}{\left|\vec{n}
+n_0 \frac{\vec{k}}{\omega }\right|}$$
and essentially we noticed that $\vec P$ can be written in terms
of a susceptivity tensor $\chi$
that can be expressed in terms of $\hat v$ as follows:

\begin{equation}
 \chi\equiv \chi(\vec n,n_0,\vec k,\omega)=  \frac{2}{M_{P}} \omega \left|\left(\vec{n}
+n_0 \frac{\vec{k}}{\omega }\right)\right|\left(n_0
+\vec{n}\cdot \frac{\vec{k}}{\omega }\right)^2\;\left[
\begin{array}{rrr}
0 & i \hat v_3 & -i \hat v_2 \\
-i \hat v_3 & 0 & i \hat v_1 \\
i \hat v_2 & -i \hat v_1 & 0
\end{array}\right]\label{eq:chi}
\end{equation}

Clearly the availability of a strict
analogy between our model and propagation in anisotropic media
is confined to the ideal case of propagation of
plane waves, since the susceptivity tensor $\chi$ which we formally
introduced depends on the wave vector $\vec k$ (so the propagation of
generic waves, spread over different wave vectors, could not be characterized
in terms of a susceptivity tensor).
And even restricting one's attention on plane waves
there are some peculiarities
 that characterize our Planck-scale deformed
propagation of electromagnetic waves, as a result of the fact that the relation between
polarization and electric field depends on $\vec k$ and $\omega$.

For a first level of characterization of these peculiarities we can formally think
of our $\hat v$ as an effective direction of anisotropy, in which case
 one obtains a close analogy between our theory and
 the established description of propagation of ordinary electromagnetic
 waves in gyrotropic
   optically active media \cite{Landau}.
   Indeed for gyrotropic media with both natural and induced optical activity,
    the polarization vector can be written as \cite{Silverman}
\begin{equation}
 \vec P= -i f \left(\hat k\times\vec E\right) -i g \left(\hat g\times \vec E\right),
\end{equation}
where $\hat g$ identifies the direction of the external field
which induces optical activity (gyrotropic axis), while $f$
and $g$ are two coefficients for the magnitude of the effect.
The case of propagation in an inactive dielectric is obtained
for $f=g=0$, while for $f\neq0,\, g=0$ one has natural optical activity,
and the case $f=0,\, g\neq 0 $ gives pure induced gyrotropy.
As shown above, in our model $\vec P$ can be written in the same form,
with   $\hat g\to \hat n$, and
\begin{eqnarray}
 f&\to& \frac{2}{M_P}|\vec k| \left(n_0+ \frac{\scal{n}{k}}{\omega}\right)^2 n_0 \\
g&\to& \frac{2}{M_P}\omega \left(n_0+ \frac{\scal{n}{k}}{\omega}\right)^2 |\vec n|.
\end{eqnarray}
So the peculiarity of our model resides in the dependence of both $f$
and $g$ on the frequency and wave vector of the wave,
and different regimes of our model end up producing effects that resemble the
ones found in different types of anisotropic materials.
For plane waves propagating with $k_\mu$ orthogonal to  $n_\alpha$
(\textit{i.e.} $\left(n_0+ \frac{\scal{n}{k}}{\omega}\right)=0$),
  both $f$ and $g$ vanish and the system behaves
 classically (inactive dielectric).
  If $|\vec n|=0$ (which, as mentioned, is the case of the original
  Myers-Pospelov model \cite{Myers2003}) then $g=0$,
  and the system behaves like a naturally optically active medium.
  In the opposite limit, $n_0=0$,
  one has $f=0$, \textit{i.e.} a medium with pure induced gyrotropy.

\section{Dispersion, birefringence and longitudinal component}
Already on the basis of established features for the original
Myers-Pospelov model (our case $|\vec n|=0$) we must expect that the speed
of propagation of our Planck-scale deformed electromagnetic waves
should depend on their wavelength and on polarization.
For the more general case $|\vec n| \neq 0$ we shall also characterize a dependence
of these effects on the angle formed by the wave vector and the vector $\vec n$.
Moreover, while the field still has only two degreees of freedom,
a longitudinal component will in general be present: the presence of
a longitudinal component is prevented when both gauge invariange and Lorentz symmetry
hold, but our framework (while being gauge invariant) clearly breaks Lorentz symmetry.
There was no longitudinal component for solutions
of the original
Myers-Pospelov model,
but only in some sense accidentally,
as an indirect result of the adopted simplification of spatial isotropy ($|\vec n|=0$).

We shall characterize these features,
at leading order in $M_P^{-1}$,
by examining
the equation of motion
for the electric field in momentum space
that is obtained from our modified Maxwell equations (\ref{eq:maxwell}):

\begin{equation}\label{eq:campo-elettrico}
 -\vec k  (\vec k \cdot \vec E)+\vec k^2\vec E=  \omega^2 \vec E +\frac{2 i}{M_{P}} \omega (n_0\omega
 + \scal{n}{k} )^2  \left|\vec{n}+n_0 \frac{\vec{k}}{\omega }\right| (\hat v \times  \vec E) ~.
\end{equation}
In particular, from this one easily infers that at leading order
the two on-shell conditions (we have indeed birefringence)
depend on $\vec k$ and $n_\alpha$ as follows:
\begin{equation}\label{eq:disp.rel}
 \omega \simeq |\vec k|\pm \frac{1 }{M_{P}} |\vec k|^2 \left( n_0
 + \frac{\scal{n}{k}}{|\vec k|}\right)^3.
\end{equation}
where the sign choice $\pm$ codifies the difference between the two on-shell conditions.

\subsection{Restricting to the spatially-isotropic case}
For the case $|\vec n|=0$ ({\it i.e.} $n_\alpha=(n_0,0,0,0)$) Eq.~(\ref{eq:disp.rel})
of course reproduces the dispersion relation
originally obtained by Myers and Pospelov:
\begin{equation}
 \omega \simeq |\vec k|\pm \frac{1 }{M_{P}} |\vec k|^2 \left( n_0\right)^3
 ~.
\end{equation}
And from Eq.~(\ref{eq:campo-elettrico}) one then easily infers that
the ``normalized field eigenstates" (waves ``on shell" with intensity 1)
 are circularly polarized:
\begin{equation}
\vec E_{\pm} =\frac{1}{\sqrt{2}} \left(
\begin{array}{c}
 0 \\
 \pm i  \\
1
\end{array}
\right)
\end{equation}
where we are using three-dimensional Jones-vector
notation\footnote{In the notation of Jones three-dimensional
vectors the field $\vec E(x,t)=Re\left[\left(E_x \hat x+E_y \hat y
+E_z\hat z  \right)e^{i(\vec k\cdot x-\omega t)}\right]$,
with $E_x,E_y,E_z$ complex numbers is represented as $\left(\begin{array}{c}
 E_x \\
  E_y \\
 E_z
\end{array}\right)$.} and we are assuming that the field propagates along the $\hat x$ direction.
As mentioned, these characteristics of on-shell waves in our framework
establish an analogy with the case of ordinary electromagnetic
plane waves propagating in a naturally optically active material.

\subsection{Spatial anisotropy in the case with no time component for the symmetry-breaking vector}
It is valuable to first compare the Myers-Pospelov/spatially-isotropic case
to the opposite regime $n_0=0$, $n_\alpha=(0,n_x,n_y,n_z)$.
In this case, $\vec n$, the spatial part of $n_\alpha$,
plays a role that is closely analogous to the role of
the gyrotropic axis for ordinary propagation
in crystals. The dispersion relation takes the form
\begin{equation}\label{eq:spatial-disp}
 \omega \simeq |\vec k|\pm \frac{1 }{M_{P}} |\vec k|^2 \left(  \frac{\scal{n}{k}}{|\vec k|}\right)^3~,
\end{equation}
so that evidently there is a strong dependence of dispersion on the angle between
the wave vector and the spatial part of symmetry-breaking vector.
For waves propagating in a direction
orthogonal to $\vec n$
the dispersion is completely absent (no difference from the undeformed theory),
while of course the dispersion reaches its maximum magnitude for
fields propagating along the $\hat n$ direction.
These two limiting cases, ${\vec k} \cdot \vec n = 0$
and ${\vec k} \times \vec n = 0$, are also peculiar in that for them
the field does not acquire a longitudinal component,
but a (ultrasmall, ``Planck-scale suppressed", but nonzero)
 longitudinal component is present in all other cases.

We find convenient to describe the field eigenstates in a orthonormal basis
that takes into account the relative orientation of the vectors $\vec k$ and $\vec n$:
\begin{equation}\label{eq:base}
 \left\{\frac{\vec k}{|\vec k|},\frac{ \hat n\times \vec k}{\sqrt{k^2
 -(\hat n\cdot \vec k)^2}}, \frac{-\vec k (\vec k\cdot \hat n)
+\hat n |\vec k|^2}{|\vec k| \sqrt{k^2-(\hat n\cdot \vec k)^2}}\right\}~.
\end{equation}
By adopting this basis we have that the first component of the field is longitudinal,
while the other two components lie in the plane orthogonal to the propagation direction.
And from Eq.~(\ref{eq:campo-elettrico}) one then easily finds that,
for a generic wave vector $\vec k$, in this basis

\begin{equation}
\vec E_{\pm} = \left(
\begin{array}{c}
\mp \frac{1}{M_P}\frac{|\vec k-(\vec k\cdot \hat n)\hat n| \,(\vec k\cdot \vec n)^2 |\vec n|}{|\vec k|^2}  \\
 \frac{\imath}{2 \sqrt{2}|\vec k|^2}\left(\pm 2 |\vec k|^2+\frac{2}{M_P}(\vec k\cdot\vec n)|\vec n|^2 \,|\vec k-(\vec k\cdot \hat n)\hat n|^2\right)  \\
 \frac{1}{2\sqrt{2} |\vec k|^2}\left(2 |\vec k|^2\mp\frac{2}{M_P}|\vec k-(\vec k\cdot \hat n)\hat n|^2\,(\vec k\cdot \vec n)\,|\vec n|^2\right)
\end{array}
\right)~.
\end{equation}

\subsection{General case}
If both the time and spatial part of the symmetry-breaking four-vector are nonzero
then we are in the most general scenario for our framework,
and of course the dispersion relation is the one of (\ref{eq:disp.rel}),
$$ \omega \simeq |\vec k|\pm \frac{1 }{M_{P}} |\vec k|^2 \left( n_0
 + \frac{\scal{n}{k}}{|\vec k|}\right)^3~.$$
Notice that  there is no dispersion and no birefringence
when $\scal{n}{k}=-n_0|\vec k|$, and from this we infer that
if $n_\alpha$ is space-like (or light-like) there must necessarily be
 a ``blind direction" (where the dispersion relation has classical form).
In the next section we shall attempt to characterize the range of directions
in the neighborhood of the blind direction where a significant suppression of
the non-classical effects occurs.

 It is also interesting to
examine the special case of waves propagating in a direction
orthogonal to $\vec n$. In this case the dispersion relation takes
Myers-Pospelov form, $\omega \simeq |\vec k|\pm M_{P}^{-1} |\vec k|^2 \left( n_0 \right)^3$,
but the field eigenstates are still different (if $|\vec n|\neq 0$)
from the ones found in the Myers-Pospelov model:
\begin{equation}
\vec E_{\pm} = \left(
\begin{array}{c}
\mp \frac{\sqrt{2}}{M_P}|\vec k||\vec n|n_0^2\\
\pm \frac{i}{\sqrt{2}}+\frac{i}{\sqrt{2}M_P}|\vec k| |\vec n|^2 n_0 \\
\frac{1}{\sqrt{2}}\mp \frac{1}{\sqrt{2}M_P}|\vec k| |\vec n|^2 n_0
\end{array}
\right)~,
\end{equation}
which is an  elliptically polarized field,
rotating in a plane not perpendicular to the propagation direction.

The solutions of the original Myers-Pospelov proposal, which we find convenient
to still write in the notation of Jones three-vectors\footnote{In the limit in
which $\vec k$ has the same direction of $\vec n$ the expression (\ref{eq:base})
is not well-defined, since the transverse components collapse. This simply means
that any pair of orthonormal vectors in the transverse plane can be used to
complete the basis and the form of \textit{e.g.} (\ref{eq:Ekparn}) is independent on this choice.}
\begin{equation}
\vec E_{\pm} = \left(
\begin{array}{c}
 0 \\
 \pm i  \\
1
\end{array}
\right) ~,\label{eq:Ekparn}
\end{equation}
emerge in our  more general framework when $\vec k$ is parallel to $\vec n$.
But the corresponding dispersion relation still carries a dependence on $|\vec n|$:
\begin{equation}
 \omega \simeq |\vec k|\pm \frac{ 1}{M_{P}} |\vec k|^2 \left( n_0
 + \epsilon_{\vec k ,\vec n}|\vec n|\right)^3,
\end{equation}
where $\epsilon_{\vec k ,\vec n}=1$ if $\vec k$ is parallel to $\vec n$
while $\epsilon_{\vec k ,\vec n}=-1$ if $\vec k$ is antiparallel to $\vec n$.
These cases in which the propagation direction is parallel (or antiparallel)
to $\vec n$ are the only ones where one finds field eingestates with
circular polarization (of course in the plane orthogonal to the propagation
direction, which is also the plane orthogonal to $\vec n$),
if $n_\alpha$ is time-like and $|\vec n|\neq 0$, $n_0 \neq 0$.
For space-like (or time-like) $n_\alpha$ there is also another
case with vanishing longitudinal component, the case of propagation directions
such that $\scal{n}{k}=-n_0|\vec k|$ (for which, as already stressed above,
 all anomalous effects disappear).

Finally let us note down the general result for the field eingestates,
for generic propagation directions such that $\vec k$
is not along the $\vec n$ direction ($|\vec k \cdot \vec n|< |\vec k| |\vec n|$),
 which in our basis (\ref{eq:base}) is
\begin{equation}
\vec E_\pm =
 \left(
\begin{array}{c}
 \mp\frac{2}{M_P}\frac{|\vec n|(\vec k\cdot \vec n+|\vec k|n_0)^2 |\vec k-(\vec k\cdot\hat n)\hat n|}{\sqrt{2}|\vec k|^2}\\
\pm \frac{i }{\sqrt{2}}+\frac{i }{\sqrt{2}M_P}\frac{|n|^2 |\vec k-(\vec k\cdot\hat n)\hat n|^2 (\vec k\cdot \vec n+|\vec k|n_0)}{|\vec k|^2}\\
\frac{1}{\sqrt{2}}\mp\frac{1}{\sqrt{2}M_P}\frac{|\vec n|^2 |\vec k-(\vec k\cdot\hat n)\hat n|^2 (\vec k\cdot \vec n+|\vec k|n_0)}{k^2}
\end{array}
\right)
\end{equation}
This general form of the eigenstates, as well as the corresponding
general form (\ref{eq:disp.rel}) of the dispersion relation,
can be naturally described in terms of the analogy  discussed in Section II
with propagation of
ordinary waves in gyrotropic media~\cite{Silverman, McClain, russo},
but of course this analogy is here of mere academic interest.

\section{Opportunities and challenges for phenomenology}
The characterization of Planck-scale-deformed electromagnetic waves given
in the previous section, is sufficient for the most used and efficacious
phenomenological analyses. For example,
 one can rely on the fact that
a wave of this sort emitted with a definite linear polarization
after long propagation times ends up loosing any trace of the original
linear polarization, because of the combined effect of dispersion and
birefringence~\cite{gleiser1,gleiser2,mattinglyLRR}
(also see Refs.~\cite{jackbire,jackbireB}).
Indeed some of the stringent bounds on the single parameter of the original
Myers-Pospelov proposal have
been established~\cite{mattinglyLRR,gleiser1,gleiser2}
by exploiting this polarization-erasing effect,
using observations of polarized light
from distant radio galaxies.

By placing the Myers-Pospelov proposal within the broader framework
of a generic symmetry-breaking four-vector $n_\alpha$ we have characterized
the possibility of effects that are in many ways similar to the ones
of the original Myers-Pospelov proposal, but with the addition
of spatial anysotropy. And the example of
observations of polarized light
from distant radio galaxies can easily illustrate how the spatial anysotropy
may reduce the strength of the implications of some observations.
In particular,
the observation of polarized light from a single distant radio galaxy
already produces definite bounds on a spatially isotropic
polarization-erasing effect, but in our more general framework, while one clearly
still finds polarization-erasing effects similar to the ones of the
original Myers-Pospelov proposal, these effects depend on the direction of propagation.
As stressed in the previous section,
in the cases with space-like (or time-like) symmetry-breaking vector
one even finds ``blind directions", {\it i.e.} propagation directions where no
polarization-erasing effect is produced.
In principle within our more general framework a single observation of
polarized light
from a distant radio galaxy can at best provide information on the relative
strength of different components of $n_\alpha$ but without setting any absolute
bound on the overall magnitude of the deformation.
More insightful bounds can be obtained by combining different observations,
associated with different directions of propagation,
but still producing results whose significance is partly weakened by
the lack of spatial isotropy, and only at the price of handling carefully
the fact that of course the components of $n_\alpha$ change in going from
one ``laboratory frame" to another (and therefore different data sets must be first rendered
comparable by mapping them all to a single reference frame).
It seems that the goal of achieving ``Planck-scale sensitivity"
for our more general framework requires us to either rely on data on a large sample
of directions of propagation or on data that characterize at once a sizeable range
of directions of propagation, as is the case for certain types of studies
done in cosmology~\cite{giuliaJCAP}.

We are clearly advocating the development of a rather challenging phenomenological
programme, but at least according to some perspectives from the theory side
it may well be worth the effort. There are some
intrisic reasons of interest in both the original
Myers-Pospelov proposal and the generalization we here advocated,
and, perhaps even more significantly, these studies may play the role
of a testing ground for the maturity of the techniques developed
in ``Quantum Gravity Phenomenology"~\cite{gacLRR}. This branch of phenomenology, which only
gained some momentum over the last decade, is only meaningful if
it is able to provide intelligible
indications\footnote{Readers who have followed
the recent literature on modifications of Maxwell equation (with or without quantum gravity)
will notice that works inspired by the Myers-Pospelov proposal
adopt a perspective that is complementary to the one adopted for example in
the study of the ``Standard Model Extension" programme~\cite{smeOLD,smeNEW}.
Whereas on one side one seeks specific proposals, which are then
forcefully scrutinized conceptually
and phenomenologically, on the other side one aims at parametrizing all possible departures
from Lorentz symmetry and focuses on characterizing the contours of the region of the
resulting huge parameter space that are consistent with observations.
The original formulation of the ``Standard Model Extension"~\cite{smeOLD}
was devised so that the Lagrangian density would involve
 only terms of dimension 4 or lower, and therefore did not include
 the $ \Delta \mathcal L_{QG}$ of Eq.~(\ref{eq:lagrangian}). In recent years
 a generalization of the ``Standard Model Extension" has been adopted
 (see, {\it e.g.}, Ref.~\cite{smeNEW}), allowing also for the presence
 of terms of dimension 5, like $ \Delta \mathcal L_{QG}$, and 6.
 But because of the difference in perspectives the results we here report
 have not been sought in the development of the ``Standard Model Extension".}
to the theory side of research
of the quantum-gravity problem. It is in light of these considerations that we
argue that it might be appropriate to devote a vigorous effort
toward the objective of excluding with Planck-scale sensitivity all modifications
of electromagnetism that violate Lorentz symmetry via a symmetry-breaking four vector
and are formulated as dimension-five corrections within effective low-energy field theory.

Since it appears that from the viewpoint of phenomenology the most challenging
aspect of our framework is indeed the spatial anysotropy and particularly the
mentioned ``blind directions", in figures 1 and 2 we provide a more quantitative characterization
of these features.
For definiteness in these figures we took $n_0>0$.
And we only considered cases with $n^\alpha$ spacelike ($|\vec n|>n_0$),
since these clearly are the cases for which the previous literature (focused on
the Myers-Pospelov case $n_\alpha = (n_0,0,0,0)$)) informs less reliably our intuition.

\begin{figure}[htb]
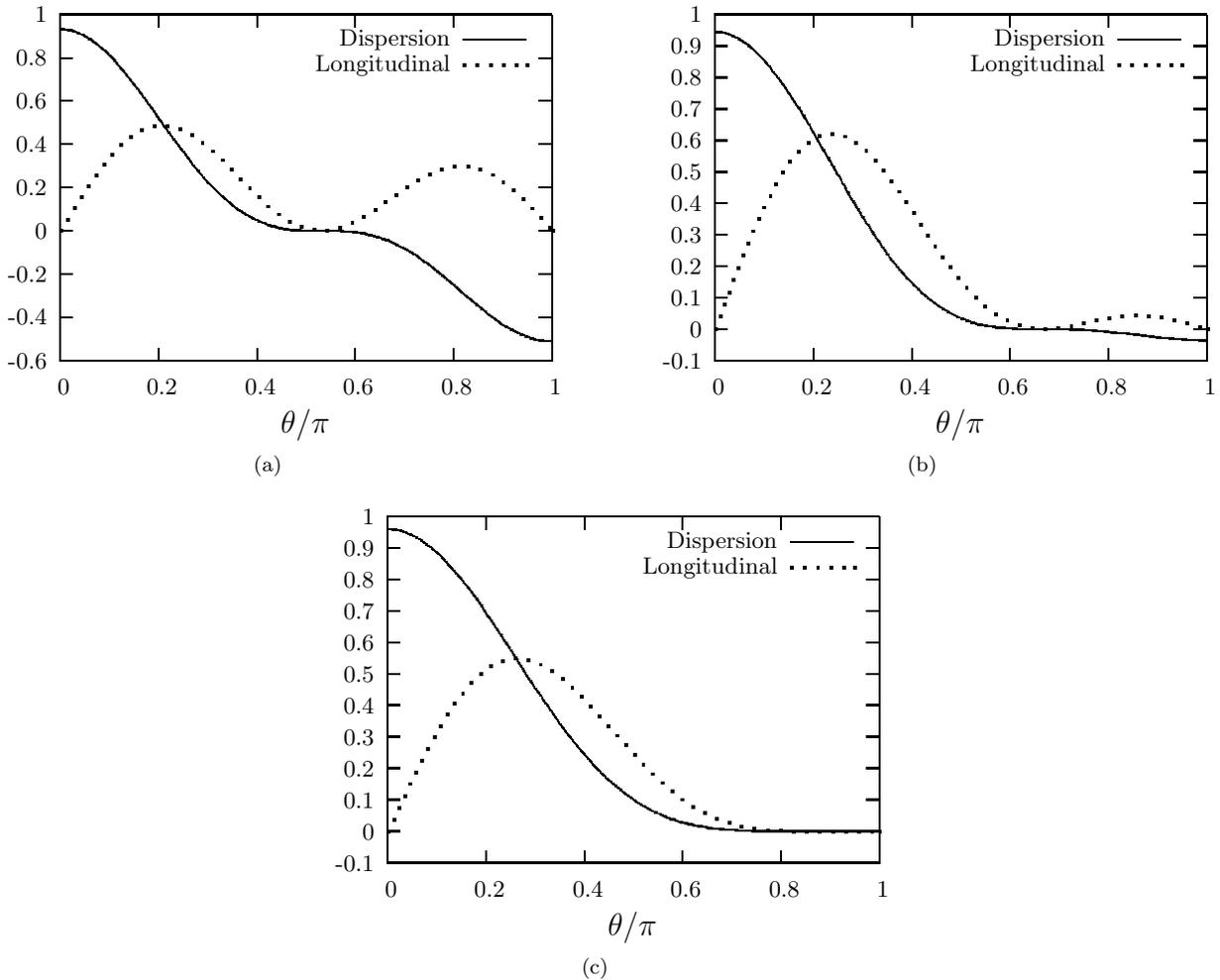

\begin{center}
\subfigure[]{\input{plots.tex}}
\subfigure[]{\input{plots2.tex}}
\subfigure[]{\input{plots3.tex}}
\caption{Behaviour of the longitudinal component of the field (dotted line),
 and of the nonclassical part of the dispersion law (continuous line) as
 functions of $\frac{\theta}{\pi}$. For the behaviour of the dispersive effects
 we simply show (up to an irrelevant overall factor introduced for visibility)
 the function $(n_0+|\vec n|\cos (\theta))^3$,
which indeed gives the dependence of these effect on the propagation direction,
and similarly for
 the longitudinal component we show (up to another irrelevant factor
 introduced for visibility) $(n_0+|\vec n|\cos (\theta))^2 \sin (\theta)$,
 which indeed gives the dependence of the longitudinal component on the propagation direction.
 For panel (a) we took $|\vec n|=1$,$n_0=0.1$; for panel (b) we took
 $|\vec n|=1$,$n_0=0.5$,
 and for panel (c) we took $|\vec n|=1$,$n_0=0.9$.}

\end{center}

\end{figure}
\begin{figure}[ht]
\begin{center}

\input{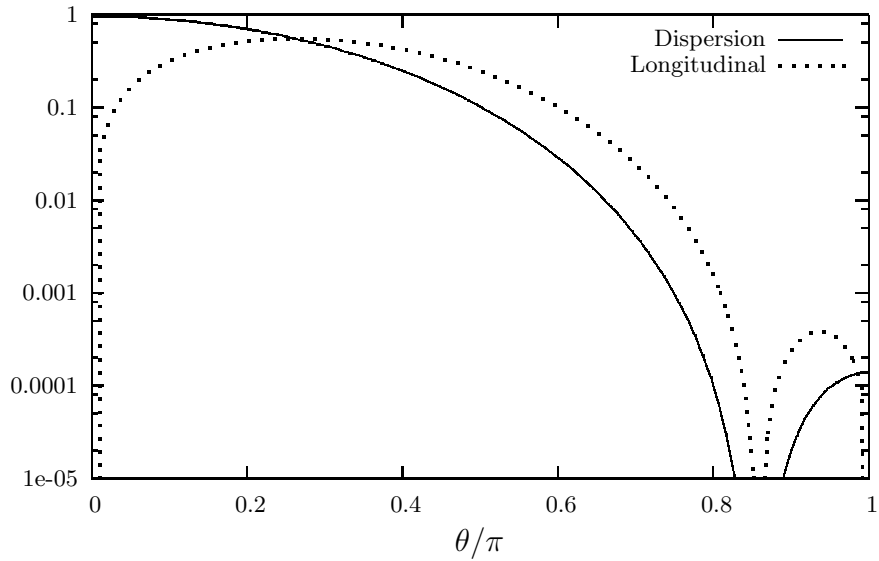}
\caption{Here we show that same case already shown in figure 1.c, but in a logarithmic plot
which allows to better appreciate the sizeable ``partial blindness" present
between $\theta \simeq 0.8 \pi$ to $\theta \simeq \pi$.
[Notice that, this being a logarithmic plot, we characterize the dispersive effects
by the absolute value of $(n_0+|\vec n|\cos (\theta))^3$,
whereas figure 1.c showed also the behaviour of the sign
of $|(n_0+|\vec n|\cos (\theta))^3|$.]}

\end{center}

\end{figure}

The figures highlight the key role played by the angle $\theta$ between $\vec n$,
the spatial part of the symmetry-breaking vector,
and the direction of propagation  ($cos \theta \equiv \frac{\vec n \cdot \vec k}{|\vec n||\vec k|}$),
and they
mainly intend to characterize the behaviour of the correction to the
dispersion relation (and associated birefringence), but also describe the behaviour of the
longitudinal component of the field.
From the figures one easily recognizes several characteristic features, some of which we had already
pointed out in our discussion of some relevant formulas:
\begin{itemize}
             \item the longitudinal component vanishes both for $\theta =0$ and for $\theta = \pi$;
             \item the magnitude of the dispersive effects is greatest for $\theta =0$ (would have been
              greatest for $\theta = \pi$ if we had chosen $n_0 <0$);
         \item in all figures one clearly notices the ``blind" value
         of $\theta$, $\theta=\theta_0$, with $\theta_0$ such that $n_0 + |\vec n| \cos \theta_0 =0$,
         where both the longitudinal component and the dispersive effects vanish;
         \item the smallness of the dispersive effects persists for a sizeable range
         of values of $\theta$ in some neighborhood of $\theta=\theta_0$.
        \end{itemize}

This last point was particularly surprising for us: at the qualitative level we expected of course
that the dispersive effects would be small in some neighborhood of the blind direction,
but somehow we envisaged this neighborhood would be very small. Instead one typically finds
sizeable regions of ``partial (but significant) blindness".
In order to render this feature more visible in figure 2 we show in logarhitmic scale
the same case
already shown in figure 1.c,
with $n_0/|\vec n| = 0.9$ and blind direction $\theta_0 \simeq 0.86 \pi$.
Comparable ``blindness features"
are found for all cases with space-like $n_\alpha$.
It is noteworthy that in figure 1.c one sees
 a rather persistent suppression of the dispersive effects by more than 4 orders of magnitude,
 all the way from $\theta \simeq 0.8 \pi$ to $\theta \simeq \pi$.

\section{Closing remarks}
The fast pace of improvement of the phenomenology
of the Myers-Pospelov proposal~\cite{Myers2003}
exploited
the spatial isotropy regained by the {\it ad hoc}
choice $n_\alpha = (n_0,0,0,0)$.
It should be noticed that the analysis we reported here
is in principle relevant even for the case of timelike $n_\alpha$,
where $n_\alpha = (n_0,0,0,0)$ is possible in one class of frames:
for frames boosted with respect to a frame with $n_\alpha = (n_0,0,0,0)$
there would of course be a spatial component for $n_\alpha$.
There is a recent literature~\cite{gacdsr,leedsr,kowadsr} on frameworks
that could implement observer-independent
departures from Lorentz symmetry,
but this is clearly not the case of the Myers-Pospelov setup.
 Phenomenologists who have analyzed the
 Myers-Pospelov proposal are well aware of this frame dependence,
 and they have neglected it only in light of the fact that
 the different ``laboratory frames" where the data were being collected
 are connected by relatively small boosts.
 For the case of time-like $n_\alpha$ our analysis is therefore at least
 valuable in as much as it allows to actually estimate the size
 of corrections that are being neglected by assuming
 that $n_\alpha = (n_0,0,0,0)$ in all of these laboratory frames.

Of course, the most intriguing part of our findings concerns
the case of space-like $n_\alpha$, which had not been considered
in previous works inspired by the Myers-Pospelov proposal.
Of course, for space-like $n_\alpha$ one could easily imagine that there
would be sizeable anysotropy, but it might have been hard to imagine that,
for example, a reduction of anomalous effects by 4 orders of magnitude (``partial blindness")
could persist for ranges of directions of propagation (with respect to the direction of $\vec n$)
as large as shown in the previous section.
At least at the preliminary level of analysis we here offered it appears that
these features might reduce significantly the bound-setting power
of most types of observations in astrophysics, which essentially probe a narrow range
of directions of propagation. And instead the type of studies in cosmology that
can be used~\cite{giuliaJCAP} to investigate anomalous laws of of propagation
naturally involve a large range of directions of propagation. So, amusingly, the study
of our framework with space-like $n_\alpha$ might be a first instance with
a level playing field between astrophysics and cosmology in the study
of Planck-scale departures from Lorentz symmetry,
in which instead so far astrophysics was clearly in the lead~\cite{gacLRR}.

\section*{Acknowledgments}
G.~G. is supported by ASI contract I/016/07/0 "COFIS".
G.~A.-C. is supported in part by grant RFP2-08-02 from The Foundational
Questions Institute (fqxi.org).
We gratefully acknowledge conversations with R.~Durrer and S.~Liberati.

\end{document}